\documentclass[journal]{IEEEtran}
\usepackage{cite}
\usepackage{graphicx}
\usepackage{float}
\usepackage{amsmath}
\usepackage{caption}
\usepackage{subcaption}
\usepackage{balance}

\usepackage{ifthen}
\usepackage{amssymb}
\usepackage{xcolor}

\newboolean{showcomments}
\setboolean{showcomments}{true}

\makeatletter
\newcommand{\mynote}[3]{%
  \ifthenelse{\boolean{showcomments}}{%
   \fbox{\bfseries\sffamily\scriptsize#1}%
   {\small$\blacktriangleright$\textsf{\emph{\color{#3}{#2}}}$\blacktriangleleft$}}%
  {%
   \@bsphack
   \@esphack
  }%
}
\makeatother

  \renewenvironment{thebibliography}[1]{%
    \begin{oldthebibliography}{#1}%
      \setlength{\parskip}{0ex}%
      \setlength{\itemsep}{0ex}%
  }%
  {%
    \end{oldthebibliography}%
  }

\begin{document}

\title{Pinning Fault Mode Modeling for DWM Shifting}

\author{Kawsher Roxy$^*$, Stephen Longofono$^\dagger$, Sebastien Olliver$^\dagger$, Sanjukta Bhanja$^*$, and Alex K. Jones$^\dagger$\\
$^*$University of South Florida $^\dagger$University of Pittsburgh\vspace{-.1in}}%

\maketitle

\begin{abstract}
Extreme scaling for purposes of achieving higher density and lower energy continues to increase the probability of memory faults.  For domain wall (DW) memories, 
 misalignment faults arise when aligning domains with access points.  
A previously understudied type of shifting fault, a \textit{pinning fault} 
may occur due to non-uniform pinning potential distribution caused by notches with fabrication imperfections. This non-uniformity can pin a wall during current-induced DW motion. This paper provides a model of geometric variations varying width, depth, and curvature variations of a notch, their impacts on the critical shift current, and a study of the resulting impact on fault rates of DW memory systems. An increase in the effective critical shift current due to 5\% variation predicts a pinning fault rate on the order of $10^{-8}$ per shift, which results in a mean-time-to-failure of circa 2s for a DW memory system.
\end{abstract}


\IEEEpeerreviewmaketitle

\section{Introduction}

Domain wall memories (DWMs) are formed from ferromagnetic nanowires.  These nanowires extend the free layer of the magnetic tunnel junction (MTJ) used to form spin-transfer-torque magnetic random access memories (STT-MRAM).  Each DWM nanowire may store multiple data-bits, \textit{e.g.,} 32-512 bits, separated by fabricated notches to form magnetic domains. Domains store these bit values through magnetic polarization.  Between adjacent domains storing complimentary bits, a mobile \textit{domain wall} (DW) is formed to balance the exchange and
anisotropic energies~\cite{book94}. Spin-polarized current drives propagation of magnetic domains for stable domains and controlled DW motion. Fast, low-energy access and high endurance makes DWM a promising data storage device~\cite{book95} and creates potential for logic implementations~\cite{book96}. 

To achieve improved density, data shifting is necessary to align data with access points.  
Thus, \textcolor{black}{shift fault characterization from process variation must be characterized to allow DWM designs to appropriately tradeoff efficiency and reliability.}  
There has been significant effort on power reduction and speed improvement~\cite{TapeCache,DWM_Tapestri,a,b} to optimize DWM shifting overhead; relatively fewer efforts~\cite{hifi,ollivier2019dsn} focus on shift \textit{reliability} due in part to insufficient error-modeling, \textcolor{black}{which we address here}. 

DWs are held at \textit{pinning sites} from intentionally engineered notches at regular intervals to ensure controlled shifting and alignment.  The notches create a potential well that must be overcome to allow DW motion.  \textcolor{black}{While many notch shapes have been explored, triangular notches have become the standard choice.}  Triangular notches are popular as they minimize shifting energy while being fairly invariant to thermal effects. Slight over- or under-shifted DWs tend to \emph{relax} toward pinning sites.  Notch deformation creates non-uniform pinning strength.   Sufficiently deformed notches can keep a DW pinned when applying the rated shift current. 

In this paper, we are the first to advance error modeling for DWM by demonstrating and characterizing \textit{uncontrolled nanowire pinning} during DW motion.  We model geometric notch variation impact of width, depth, and \textit{curvature} by 10\% on the critical shift current as representative fabrication process variation.  
Our model demonstrates that runtime \textit{pinning faults} occur at a rate of circa $1.6\cdot 10^{-8}$.  \textcolor{black}{Next we discuss the system-level motivation and need for development of this fault model.}

\section{Motivation}

\begin{figure}[tp]
 \centering
\includegraphics[width=\columnwidth]{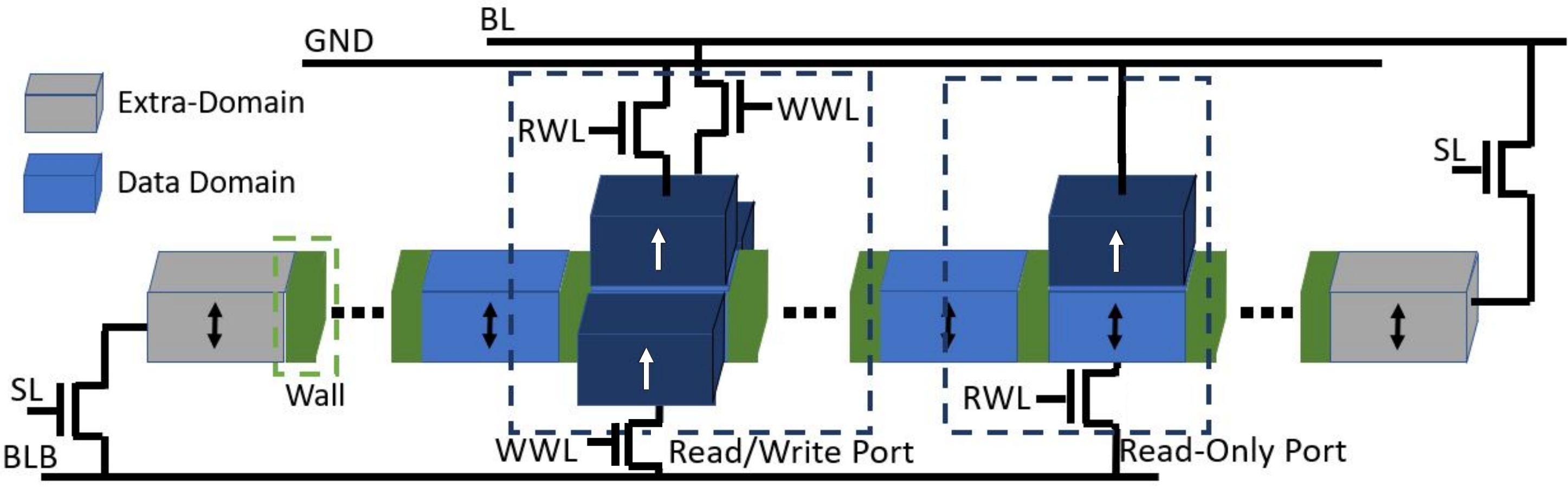}
\caption{Anatomy of a DWM nanowire~\cite{DWM_Tapestri}.}
\label{DWMzoom}
\label{fig:nanowire}
\vspace{-.2in}
\end{figure}

\textcolor{black}{An example of a planar (2D) DWM nanowire with shift write ports is shown in Fig.~\ref{fig:nanowire}~\cite{DWM_Tapestri}.  The value of each domain is determined by its polarization as illustrated by arrow direction. Access ports (read and/or write locations) are fixed elements. Thus, a domain needs to be shifted and aligned with the access port to be read or written.  This requires ``padding'' domains (shown in gray) on each side of the data domains to prevent data loss.  During shifting, a pulse is sent along the nanowire.  Once a domain is aligned with the head, an orthogonal current to the nanowire along the access port fixed layer can be used to read or write the domain like an MTJ.  Shift based write ports allow writing a domain by shifting orthogonally to the nanowire from fixed magnetization domains, in order to reduce energy due to high writing currents~\cite{DWM_Tapestri}.}

\begin{figure}[bp]
\centering
\vspace{-.2in}
\includegraphics[width=\columnwidth]{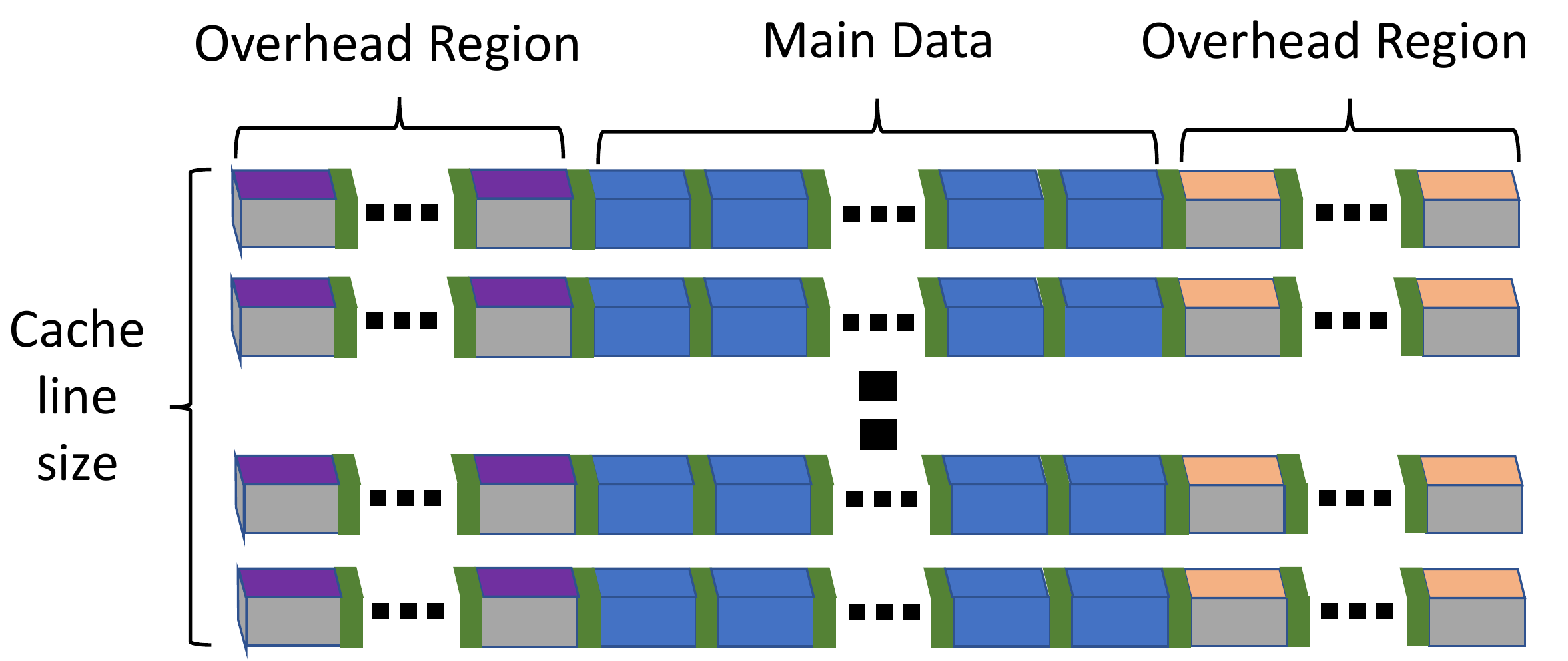}
\caption{Domain block cluster example.}
\label{BlockNano}
\end{figure}

\textcolor{black}{To build memories from nanowires, a ``bundle'' of nanowires are grouped together to enable parallel bit-wise access to each element of a memory row.  The row data width is determined by the number of nanowires in the bundle.  A \textit{domain block cluster}~\cite{TapeCache,FusedCache,xu2015multilane,Castrillon-Polyhedral,Castrillon-ShiftsReduce} (DBC), shown for a cache line granularity in Fig.~\ref{BlockNano}, contains many $\ell$ memory rows where $\ell$ is the number of data domains in each nanowire.  Each bit of the cache line is individually accessed by shifting all the nanowires together.}

\textcolor{black}{Prior work has demonstrated that misalignment (under- or over-shift) of one or more nanowires in a DBC is possible due to various factors that include current fluctuations in the system.  Such faults occur at a rate circa $10^{-5}$~\cite{hifi}.  In these cases, the entire nanowire uniformly shifts by an incorrect amount.  Error correction codes (ECC) alone are insufficient to detect and correct these faults~\cite{ollivier2019dsn} but existing schemes have been proposed to address misalignment resulting in mean-time-to-failure (MTTF) on the order of decades~\cite{hifi,ollivier2019dsn}. } 

\textcolor{black}{Uncontrolled nanowire pinning has been demonstrated in early DWM devices~\cite{parkin2014dwm,pinning_Piramanayagam} which can occur due to process variation.  In these cases different parts of the nanowire can shift \textit{different} amounts.  When DW motion pushes towards an immutable DW due to uncontrolled pinning, \textit{i.e.,} part of the nanowire shifts one position towards the pinned domain and the remaining part does not shift, a domain is lost at the pinning point.  However, if DW motion depins too easily the nanowire may shift one position towards the pinning point and the remainder of the nanowire shifts two positions, duplicating a domain.  Both of these effects are \textit{pinning faults}.}  \textcolor{black}{Either of these pinning faults puts the nanowire and DBC in an unrecoverable state that misalignment protection cannot detect, let alone correct.  
Next we discuss in detail the energies of DW motion and pinning potential in a notch.}

\begin{figure*}[ht!]
    \centering
    \begin{subfigure}[]{.4\columnwidth}
        \includegraphics[width=\textwidth]{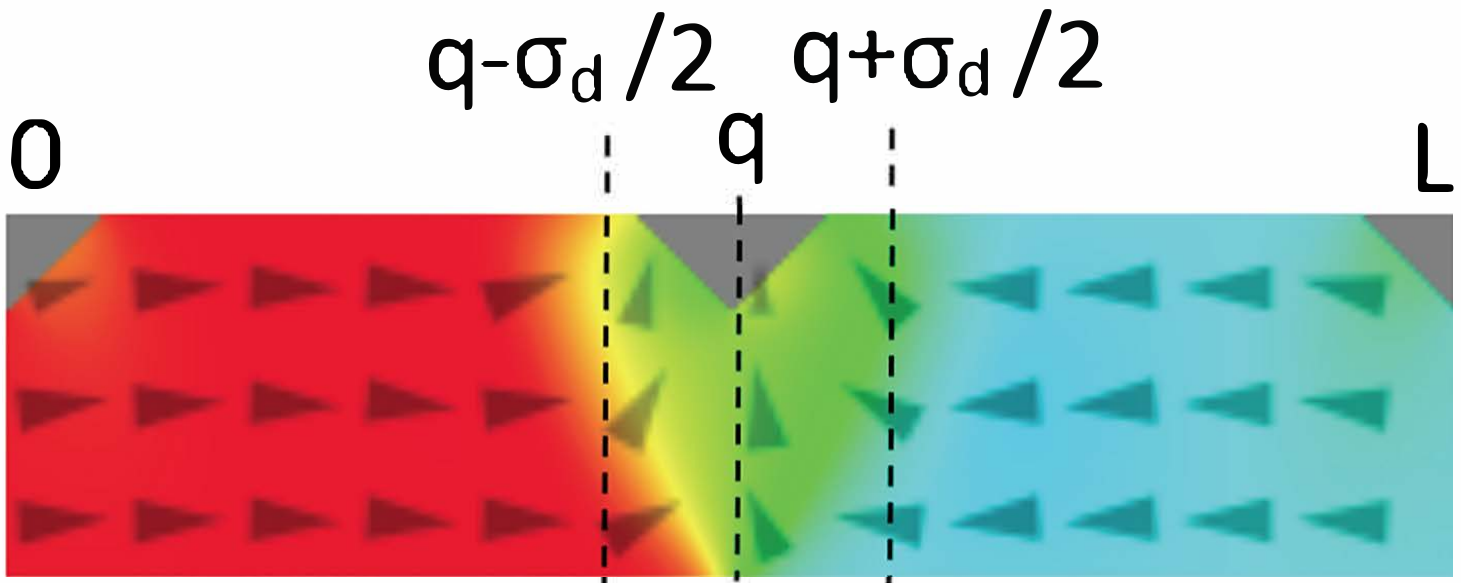}
        \caption{Pinning site}
        \label{fig:q}
    \end{subfigure}
    \begin{subfigure}[]{.55\columnwidth}
    \centering
    \includegraphics[width=\textwidth]{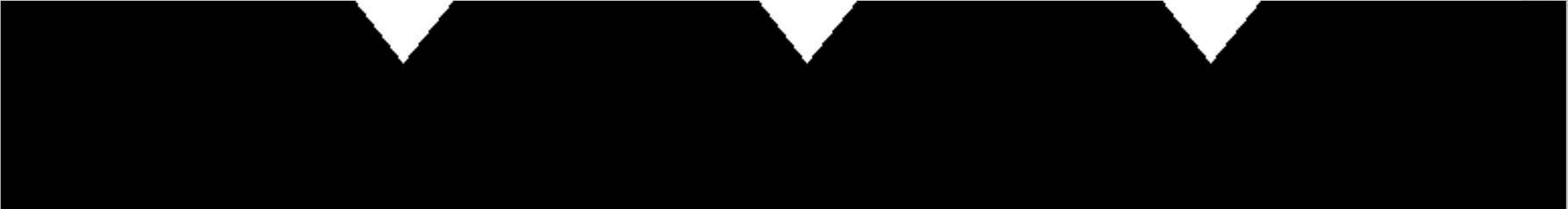}
    \caption{Nanowire with 4 domains.}
    \label{nw4}
    \end{subfigure}
    \begin{subfigure}[]{0.45\columnwidth}
    \centering
    \includegraphics[width=\textwidth, height=0.3in]{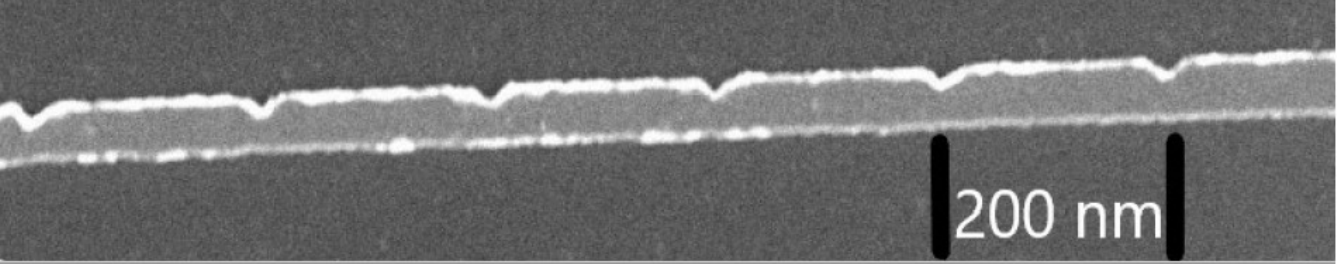}
    \caption{SEM image}
    \label{fig:sem}
    \end{subfigure}
    \begin{subfigure}[]{0.45\columnwidth}
    \centering
    \includegraphics[width=\textwidth, height=0.3in]{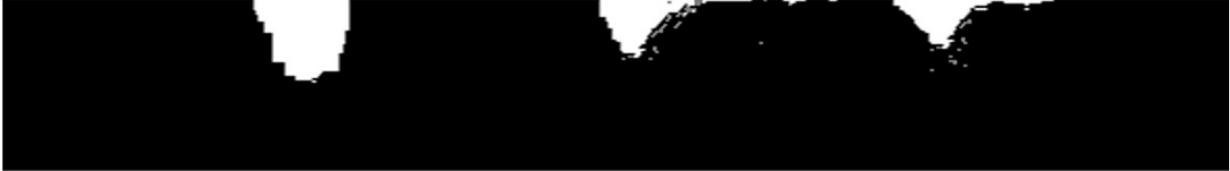}
    \caption{Bitmap from SEM image}
    \label{fig:sem-defect}
    \end{subfigure}\\
    \begin{subfigure}[]{.4\columnwidth}
    \includegraphics[height=.8in]{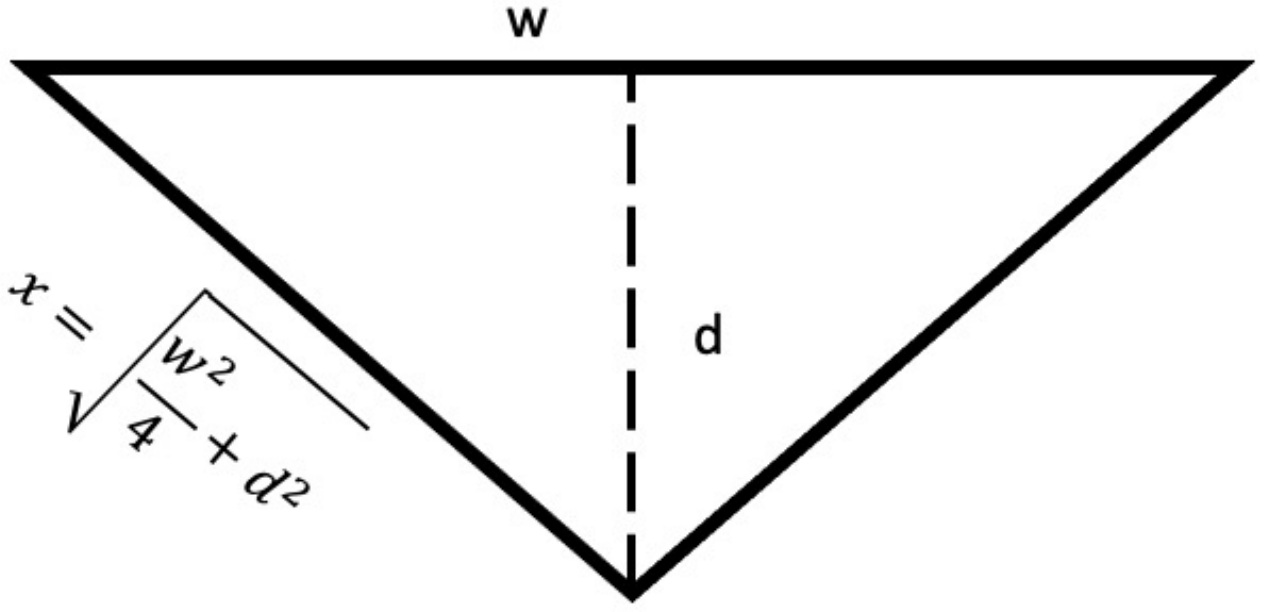}
    \caption{Notch parameters}
    \label{notch_params}
    \end{subfigure}
    \begin{subfigure}[]{0.48\columnwidth}
    \centering
    \includegraphics[height=.8in]{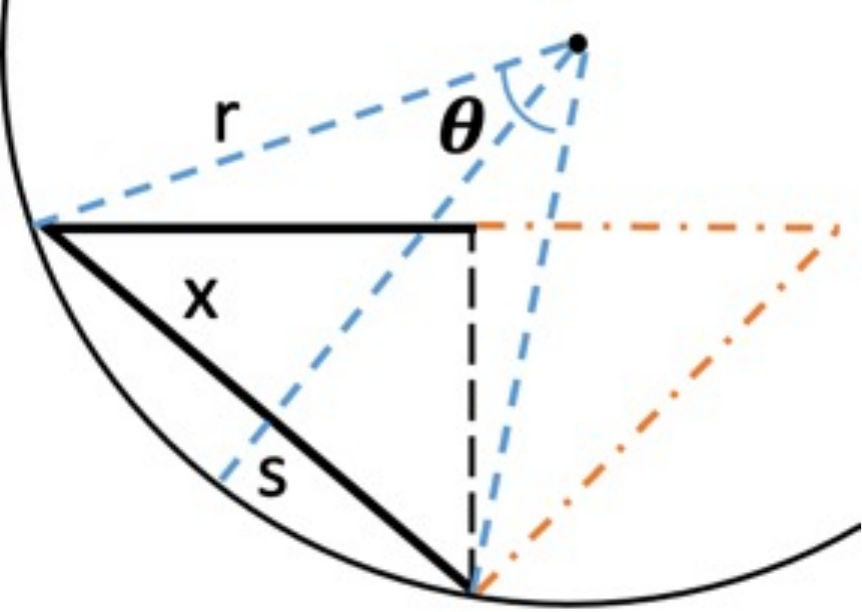}
    \caption{$r-\theta$ model: convex ($+s$)}
    \label{concave_model}
    \end{subfigure}
    \begin{subfigure}[]{0.48\columnwidth}
    \centering
    \includegraphics[height=1in]{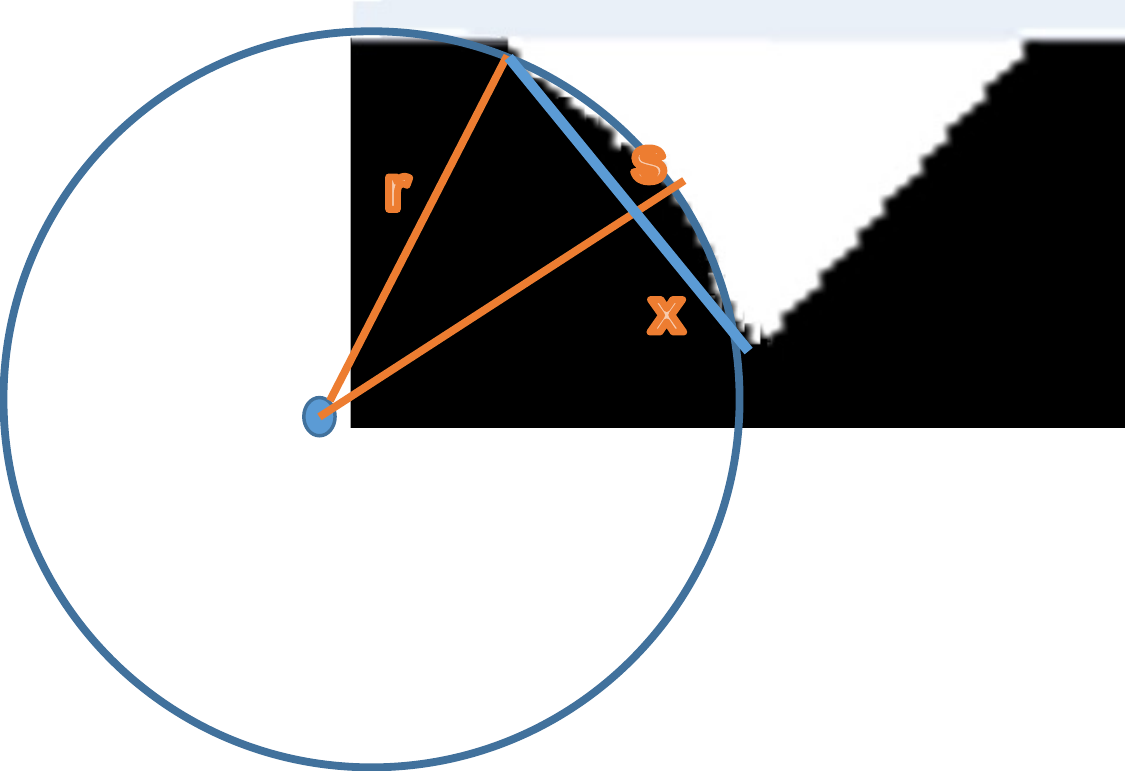}
    \caption{$r-\theta$ model: concave ($-s$)}
    \label{convex_model}
\end{subfigure}
\begin{subfigure}[]{.48\columnwidth}
\centering
\includegraphics[height=1.2in]{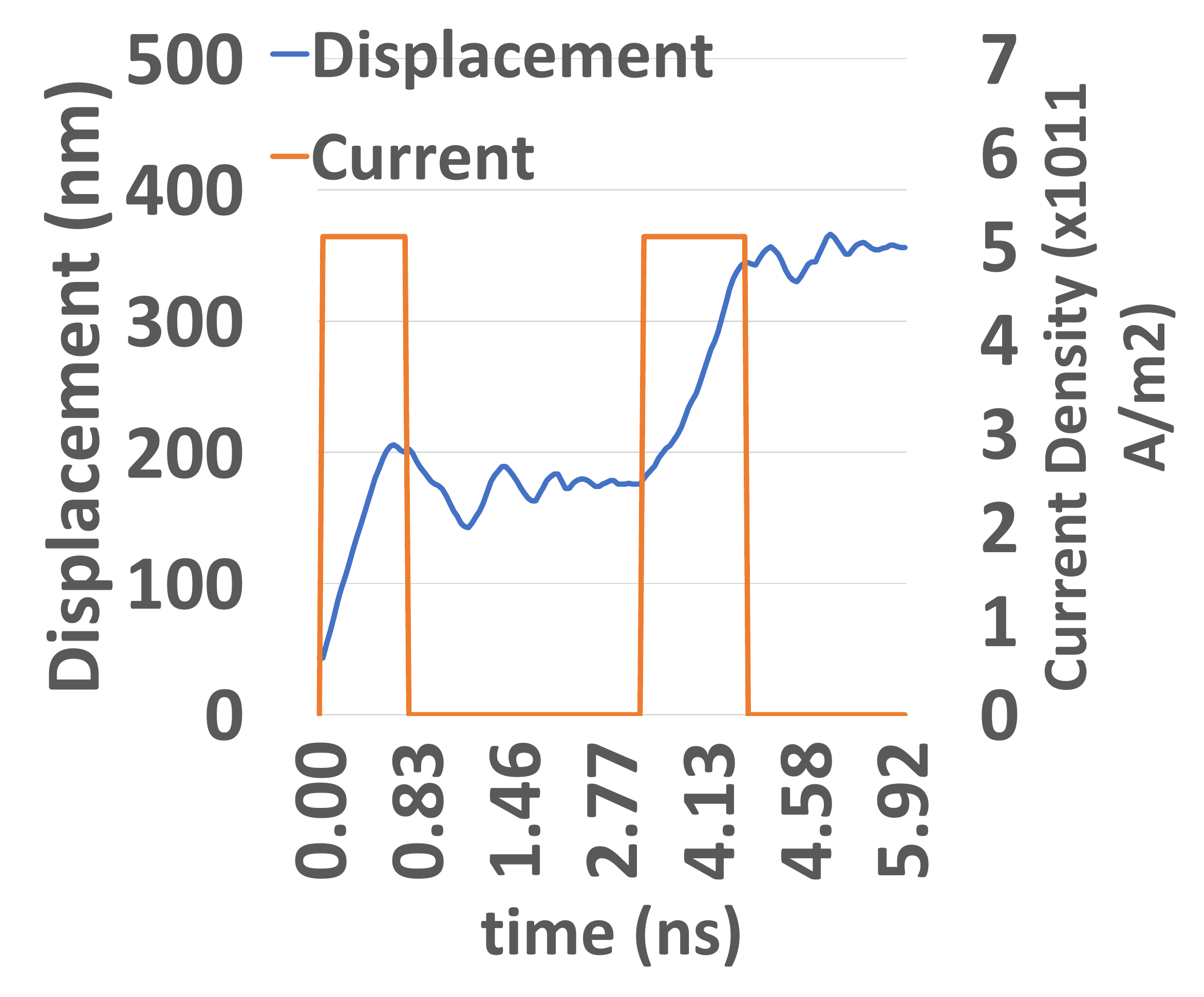}
\caption{$I_S$ pulses vs. DW movement}
\label{ch6_dw_move}
\end{subfigure}
    \caption{Process limitations leading to notch deformities, modeling the notch, and demonstration of motion in the nanowire.}
    \label{ch6_notch_curv}
    \vspace{-0.2in}
\end{figure*}

\section{Pinning of a Domain Wall}
In this section we describe the expressions that model pinning strength, current-induced depinning, the impact of process variation, and modeling notch deformations.

\vspace{-.1in}
\subsection{Pinning Potential}

The strength of a pinning site is characterized by its pinning strength or \textit{pinning potential}, which depends on the geometry of that notch. 
The pinning potential per unit area is described in Eq.~\ref{eq_pin_pot}~\cite{thomas2006oscillatory,suzuki2008analysis} as follows:
\begin{equation}
\begin{split}
    V_{pin}\!=\!\frac{2M_sE\sigma_d}{(q-q_{pin})^2}
\begin{cases}
    E\!=\!E_{pin}, q\!\in\![q_{pin}\!-\!\frac{\sigma_d}{2},q_{pin}\!+\!\frac{\sigma_d}{2}]\\
    E\!=\!0, \text{otherwise}
\end{cases}
\end{split}
\raisetag{.3in}
\label{eq_pin_pot}
\end{equation}

\noindent where \textcolor{black}{$q$ and $\sigma_d$ are the DW center and width, respectively (Fig.~\ref{fig:q})}, $q_{pin}$ is the pinning site, $V_{pin}$ is the associated pinning potential, 
and $M_s$ is the saturation magnetization of the material used.  The anisotropic and exchange energy per unit area of a notch contributes to the energy density ($E_{pin}$) of a notch~\cite{hayashi2007current}.  \textcolor{black}{$\sigma_d$ and $E_{pin}$ are represented as:
\begin{equation}
    \sigma_d=\pi\sqrt{\frac{A_{ex}}{K_u}} \; \; \; \; \; \; \; \; \; \; E_{pin}=A_{ex}\frac{\pi^2}{2\sigma_d}+\frac{\sigma_dK_u}{2}
    \label{eq:pinning-energy-density}
\end{equation}
\noindent respectively,} where $A_{ex}$ and $K_u$ are the exchange coefficient and magneto-crystalline anisotropy, respectively~\cite{sampaio2015domain,aharoni2000introduction}.  


\subsection{Current-induced Depinning of a DW}
A current pulse with adequate amplitude 
can depin a wall such that it can travel along the nanowire to the next pinning site. During current flow, the \textit{4s} conduction band electrons interact with the \textit{3d} band electrons of magnetic domains generating an exchange interaction torque~\cite{berger1984exchange}. At a domain wall the magnetization flips over a plane parallel to the wall. As a result, the conduction electrons bend at the wall while crossing it, amounting to a transfer of momentum between the passing electrons and the local magnetic moment of the wall. This transfer process is adiabatic and included in Eq.~\ref{LLG}. 
Due to the conservation of energy, the electrons exert a force on the wall, which can move it with sufficient current density~\cite{tatara2004theory}. 
The micromagnetics of shifting a DW is governed by the Landau-Lifshitz-Gilbert equation with the inclusion of current-induced torques~\cite{tatara2008microscopic} (assuming the current flows into $\vec{x}$-direction) as: 
\begin{equation}
\frac{d\vec{M}}{dt}\!=\!-\gamma\vec{M}\times\vec{H}_{eff}+\alpha\vec{M}\times\frac{d\vec{M}}{dt}-v_j\frac{\partial\vec{M}}{\partial x}+ \beta v_j\vec{M}\times\frac{\partial\vec{M}}{\partial x}
\label{LLG}
\end{equation}

\noindent where $\vec{M}$, $H_{eff}$, $\alpha$, $\gamma$, and $\beta$ are the magnetization orientation, effective field, Gilbert damping constant, gyromagnetic ratio, and non-adiabatic spin-torque coefficient, respectively. The last two terms that are functions of $\partial\vec{M}/ \partial x$ represent the current-induced torques that are responsible for DW shifting with shifting velocity of $v_j$.


\subsection{Process Variation}
As alluded to previously, triangular notches are most common in DWMs as the pinning strength of a triangular notch is sufficient to avoid depinning due to thermal perturbation but small enough to require a relatively low shift current.  An ideal triangular notch is symmetrical on the Y-axis. However, process variations can alter the symmetry, which deflects the pinning potential. A non-uniform distribution of pinning potential due to process variation impacts the shift current requirement at each notch.


Figs.~\ref{nw4}--\ref{fig:sem-defect} show the shape and geometric deformities of notches in a nanowire having domains of 200nm long and 100nm wide. An ideal notch (Fig.~\ref{nw4}) is a symmetrical triangle as shown in Fig.~\ref{notch_params} with width $w= 50$nm and depth $d=30$nm. There is geometric variation due to fabrication processes such as lithography and deposition as shown in an scanning electron microscope (SEM) image (Fig.~\ref{fig:sem}). However, notches can have depth and width variation and can be curved as opposed to having perfect line segments (Fig.~\ref{fig:sem-defect}).  We adapted the $\pm5\%$ variation of width and depth of the notches described in models in the literature~\cite{iyengar2014modeling,hayashi2007current} and expanded this to include $\pm5\%$ variation in the area increased or decreased by a curved side of the notches described next.

\subsection{Modeling Notch Deformation using Curvature}
\label{model}

To model notch deformation using curvature we interpret one side of the triangular notch as an arc of a circle that passes through the starting and ending points of the ideal notch edge.  The ideal notch edge is the chord to this arc. 
Our method to model the curvature considers both the convex (Fig.~\ref{concave_model}) and concave (Fig.~\ref{convex_model}) arcs, where the concave case is reflected about the chord. The arc length is $S=r\theta$, where $r$ is the radius of the circle and $\theta$ is the angle at the center of the circle to specify the arc.   In the case of a perfect notch, the radius $r$ increases asymptotically towards infinity and similarly $\theta$ decreases towards zero such that the arc and chord are indistinguishable. 
The degree of curvature can be quantified by the length of the 
sagitta $s$ between the curve and the ideal notch edge (chord). The sagitta represents the displacement or deflection of the farthest point of the arc from the mid point of the chord as described in Eq.~\ref{eq:sagitta-def}, where $x$ is the length of the chord.
\begin{equation}
    s=r-\sqrt{r^2-x^2/4}
    \label{eq:sagitta-def}
\end{equation}

Varying the sagitta (curvature) varies the area displaced by the notch. If the sides of a triangular notch become curved instead of straight lines, the area of the notches is increased or decreased based on the curvature type, \textit{i.e.,} convex notches have increased area and concave notches have reduced area. Similar to the method for width and depth variation, we also considered $\pm5\%$ variation of area, centered about a perfect notch, where 
the radius of the circle is infinity  which corresponds to a sagitta $s=0$.  


Recalling that Fig.~\ref{notch_params} shows our ideal triangular notch, the area of the half of the notch is $A=\frac{1}{4}wd$. Fig.~\ref{concave_model} depicts a deformed convex notch
, with the corresponding model for the concave case in Fig.~\ref{convex_model}. The increased area due to the deformed side is given by:
\begin{equation}
\label{eq:area}
    \Delta A= \frac{\theta r^2 - (r-s)\sqrt{\frac{w^2}{4}+d^2}}{2}\approx \frac{s\sqrt{\frac{w^2}{4}+d^2}}{2}
\end{equation}

For small curvature variation, $\theta$ remains sufficiently close to zero such that $S \approx x$. Thus, $\theta$ can instead be expressed as the ratio of $x$ to $r$ or as a function of $d$ and $w$ is $\sqrt{w^2/4+d^2}/r$, which reduces $\Delta A$ as shown in Eq.~\ref{eq:area}.
    
For our experiments, we change the area relative to the ideal half-notch area to form the upper bound on how $s$ is varied.


In the next section we present the results of experiments to study the impact of varying the notch width, depth, and curvature to study the impact of variation.  





\section{Results and Discussion}
In our experiments our control was a DW nanowire with triangular notches on the top edge of the nanowire with a width of 50nm and a depth of 30nm. 
Fig.~\ref{nw4} shows a conceptual layout of a typical DW nanowire segment with 4 domains. 
%
%
Our experiments vary the notch shape in terms of 
width, depth, and curvature to determine their impact of shifting behavior.
We simulated nanowires  with dimensions $3200\times 100 \times 2$nm,  
separated into 16 domains by notches every 
200nm. 
The material properties of the nanowire are listed in Table.~\ref{parameters}.

\begin{table}[h]
\centering
\vspace{-.05in}
\caption{Material properties used in simulation.}
\begin{tabular}{|l|l|l|l|l|l|}
\hline
  $A_{ex}$ (J/m) & $M_s$ (A/m) & $\alpha$ & $K_{u}$(J/m$^3$) & \textcolor{black}{a (nm)} & \textcolor{black}{$\gamma (s^{-1}$T$^{-1}$)}\\ \hline
 2.0$\times$10$^{11}$ & 6.5$\times$10$^5$ & 0.02 & $10^6$ & 0.287 & 1.76$\times$10$^{11}$\\\hline
\end{tabular}
\label{parameters}
\vspace{-.2in}
\end{table}

\begin{figure}[bp]
    \vspace{-.2in}
    \centering
    \includegraphics[width=0.75\columnwidth]{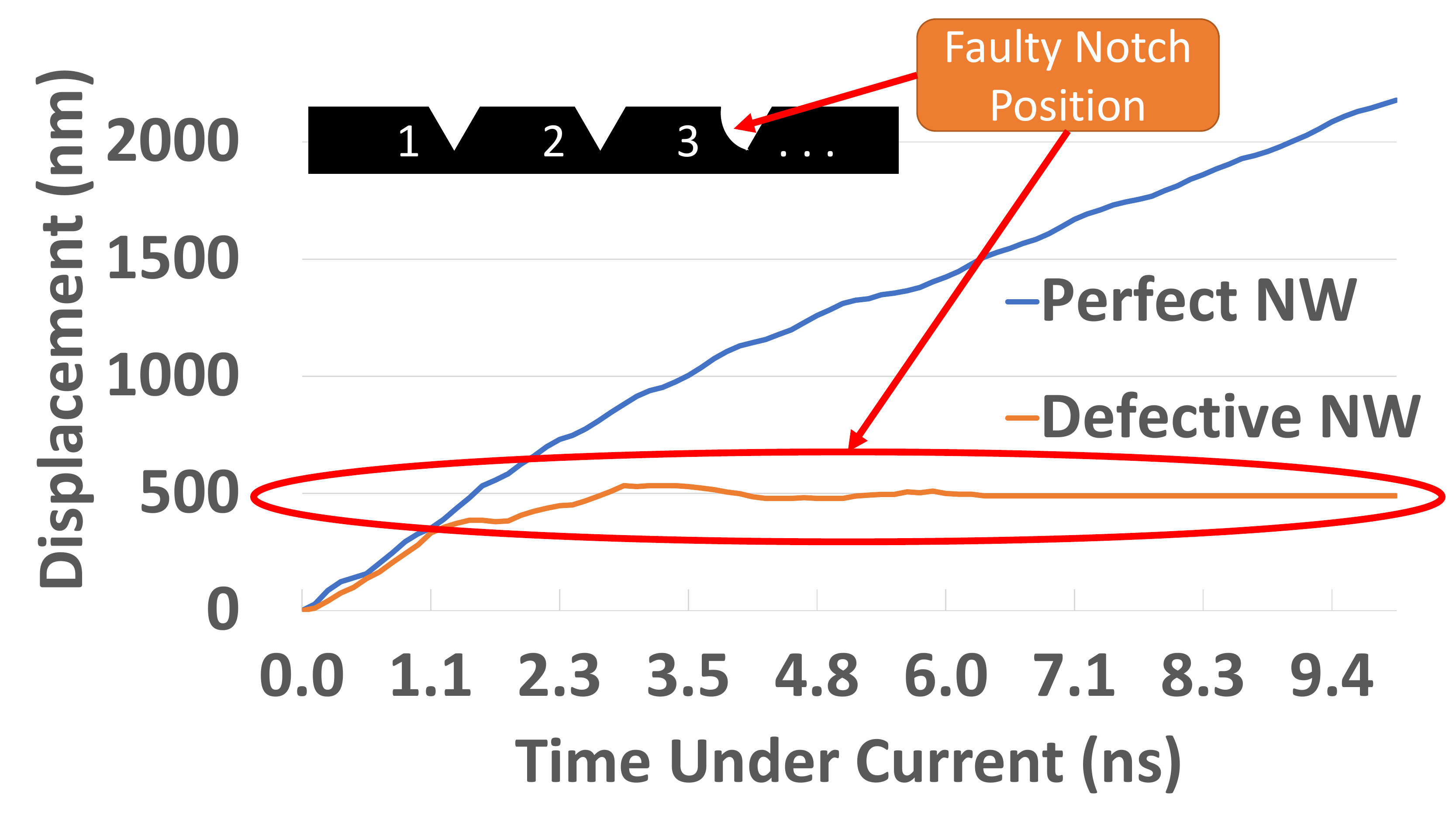}
    \caption{Domain wall stuck at a faulty notch position.}
    \label{fig:pinningDemo}
\end{figure}
\begin{figure*}[tbp]
    \centering
    \begin{subfigure}[]{0.66\columnwidth}
    \includegraphics[width=\textwidth]{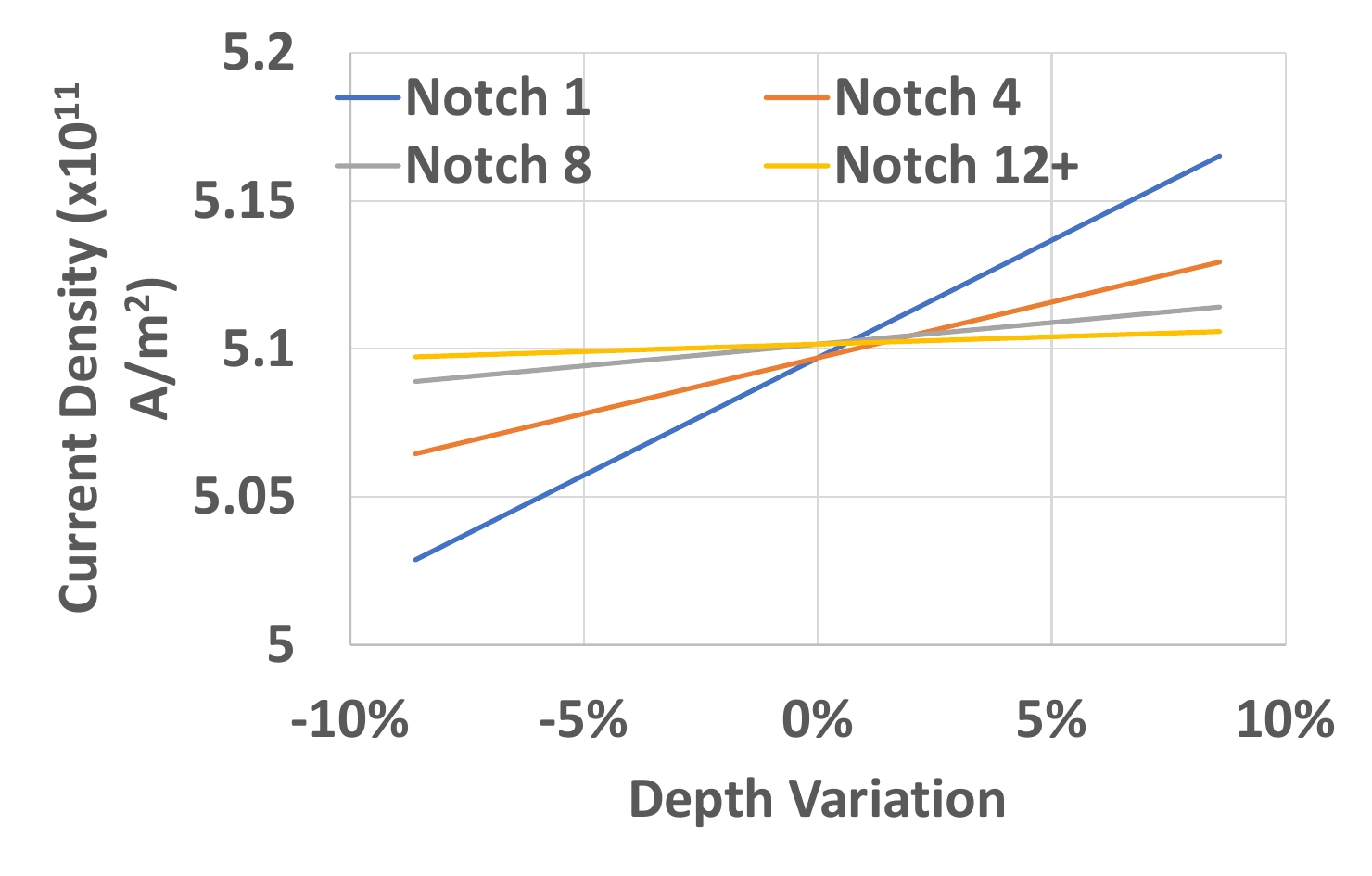}
    \caption{Width variation (CC).}
    \label{fig:width_cc_data_notches}
    \end{subfigure}
    \begin{subfigure}[]{0.66\columnwidth}
    \includegraphics[width=\textwidth]{Figures/depth-variation-akj.pdf}
    \caption{Depth variation (CC).}
    \label{fig:depth_cc_data_notches}
    \end{subfigure}
    \begin{subfigure}[]{0.58\columnwidth}
    \includegraphics[width=\textwidth]{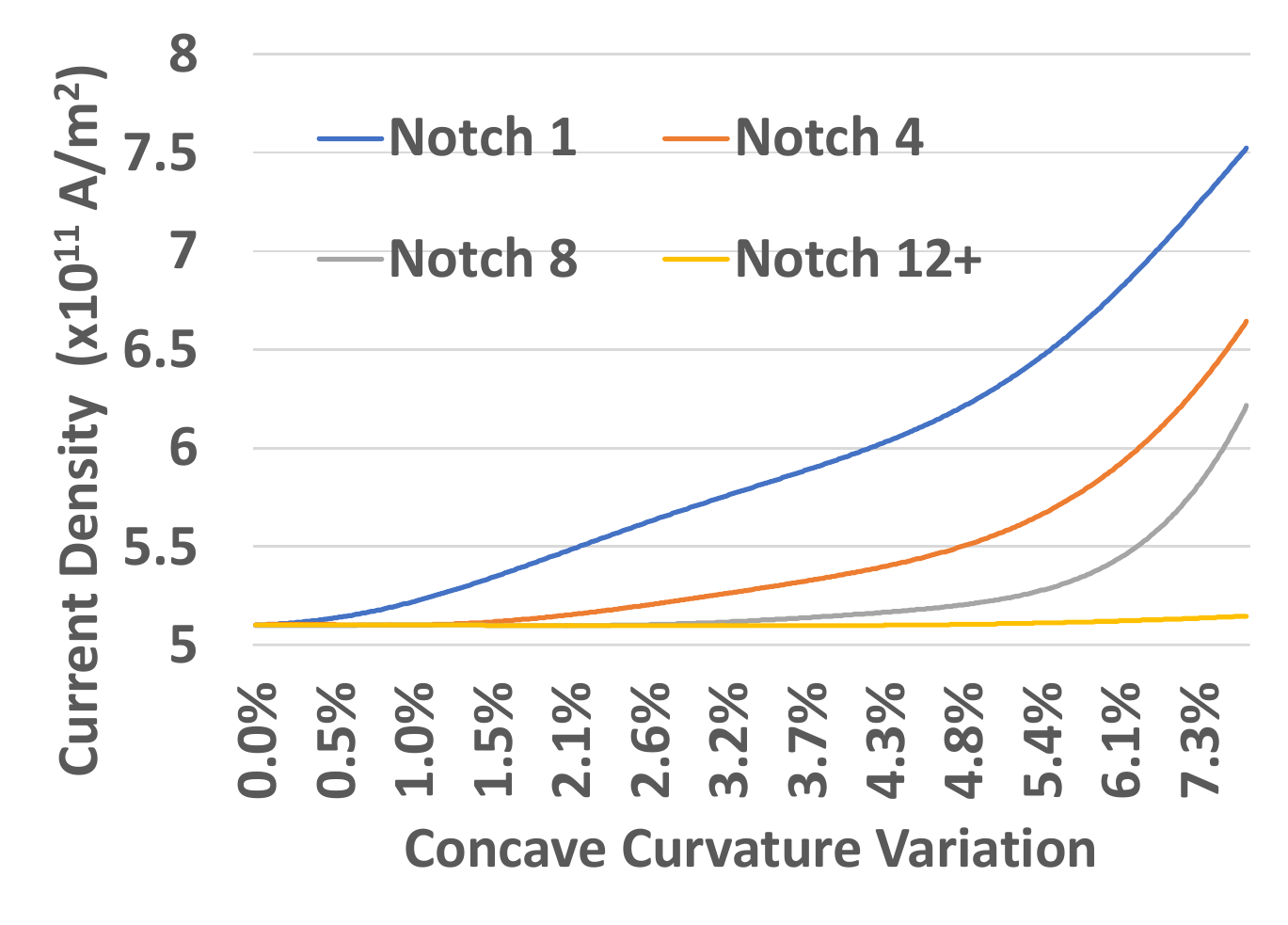}
    \caption{Convex curve variation (CC).}
    \label{fig:curv_cc_data_notch1}
    \end{subfigure}
    \\
    \begin{subfigure}[]{0.66\columnwidth}
    \includegraphics[width=\textwidth]{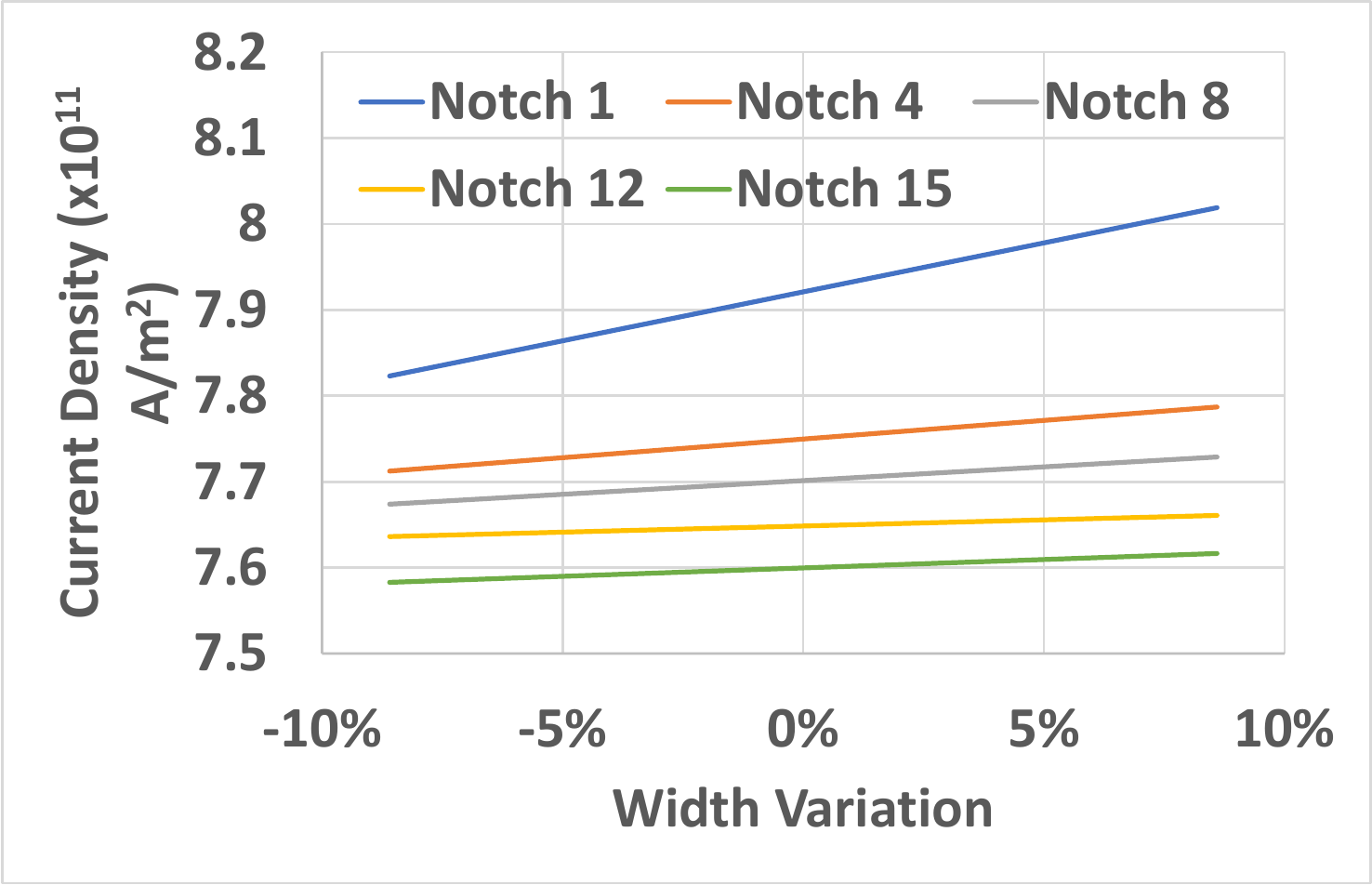}
    \caption{Width variation (UB).}
    \label{fig:width_ub_data_notches}
    \end{subfigure}
    \begin{subfigure}[]{0.66\columnwidth}
    \includegraphics[width=\textwidth]{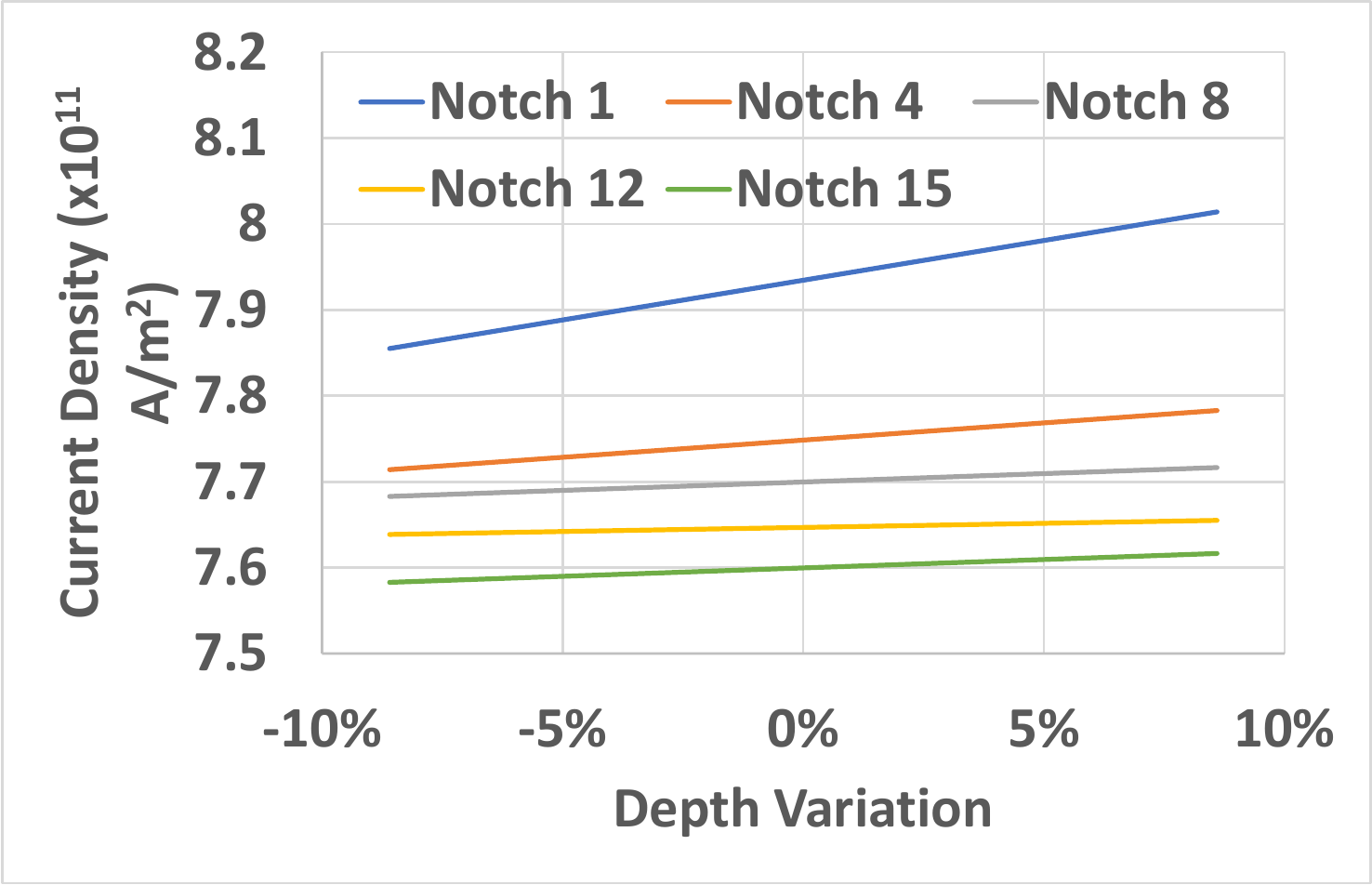}
    \caption{Depth variation (UB).}
    \label{fig:depth_ub_data_notches}
    \end{subfigure}
    \begin{subfigure}[]{0.6\columnwidth}
    \includegraphics[width=\textwidth]{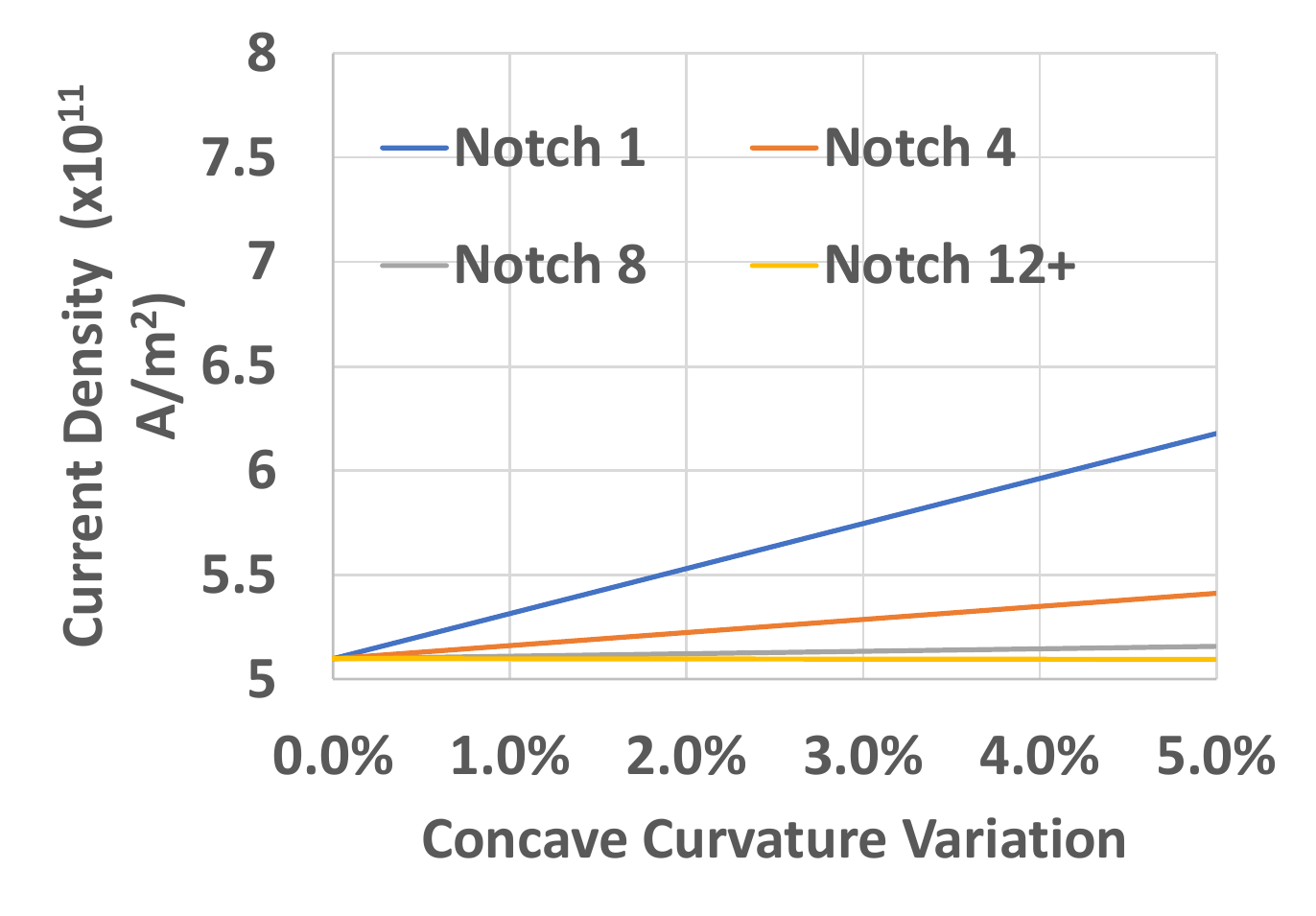}

    \caption{Linear extrapolation curvature (CC).}
    \end{subfigure}
    \caption{Change in critical shift current (CC) and associated upper bound (UB) from variation.}
    \label{ch6_curr_var}
    \vspace{-0.2in}
\end{figure*}
\subsection{Determination of Shifting Current Density Range}
We \textcolor{black}{used the open-source micromagnetic simulator, mumax3~\cite{mumax}} to calculate the critical current requirement represented in Eq.~\ref{LLG} ignoring thermal fluctuations and assuming the absence of an external magnetic field. While our experiments varied the notch shape across the variables of interest, we held the nanowire width, thickness, and length constant (no variation). We initialized the nanowire with random magnetization, and allowed the nanowire to relax. To calibrate DW movement, we applied two consecutive current pulses of 0.5ns at a 3ns interval, and tracked the DW position along the nanowire (3ns is the relaxation time after 1-bit shifting). Fig.~\ref{ch6_dw_move} shows the current pulse applied to the nanowire in orange and the subsequent x direction DW displacement in black.



We tuned the shift current to find the critical current density for shifting in this ideal scenario. A nanowire containing 16 domains with no deformed triangular notches requires a minimum current density of $5.1\times10^{11} A/m^2$ to successfully shift one bit. We also calculated the upper-bound of shift current density above which the DW shifts by two positions as $7.9\times10^{11} A/m^2$. These values serve as the control for 
comparison with deformed nanowires.

\vspace{-.1in}
\subsection{Demonstration of Pinning from Notch Deformation}
To demonstrate faulty pinning conditions, we applied the critical shift current continuously for 10ns to an ideal nanowire and a nanowire with a deformed notch in the 
3\textsuperscript{rd} and 7\textsuperscript{th} pinning sites. We introduced a DW at the first notch and measured the distance traveled in both cases. Fig.~\ref{fig:pinningDemo} demonstrates that the wall becomes stuck at the 3\textsuperscript{rd} notch, which is deformed with curvature as shown in the inset of the figure.  Domain motion stops after traveling approximately 400nm in the nanowire with the deformed notches, whereas it travels over 2000nm in the ideal nanowire. This phenomenon demonstrates an occurrence of a pinning fault. 

In the following sections we relate experiments to quantify the currents required for correct shifting when deformed notches due to perturbations in notch width, depth and curvature at different sites along the nanowire. 

\vspace{-.1in}
\subsection{Notch Width, Depth, and Curvature Variation}
To better understand the variation impact we first studied notch width and depth deviation from ideal values of 50nm and 30nm, respectively, at different pinning sites with no curvature. 
We considered 33 data-points increasing linearly in the range of 
$\pm 10\%$ for both parameters and calculated the critical currents for shifting. Each parameter was varied in isolation and the test was repeated at different notch positions.   
For example, position 1 is the notch closest to the applied shifting current. 

Fig.~\ref{fig:width_cc_data_notches} shows the impact of varying the notch width on the critical shifting current (CC). The CC varies symmetrically about the ideal CC of $5.1\times 10^{11} A/m^2$, with the largest impact at the notches closest to where shifting current enters the nanowire.  The same pattern exists for the CC when depth is varied, shown in Fig.~\ref{fig:depth_cc_data_notches}.  Considering the impact on the upper bound current for correct shifting (UB) width variation is shown in Fig.~\ref{fig:width_ub_data_notches}.  The slope varies less per unit change in notch width but notch position has a greater impact.  Again, a similar pattern holds for varying notch depth, shown in Fig.~\ref{fig:depth_ub_data_notches}.  In all cases, the trend is well-represented by a linear equation; this forms the basis for our models of the impact of notch position, width, and depth on CC and UB, respectively.


Considering approximately $10\%$ variation of area relative to the ideal notch, $\Delta A \leq0.05wd/4$. 
Using $\Delta A$ from Eq.~\ref{eq:area} we translated the curvature variation into a 
maximum sagitta length $s_{max}$.  Holding the width and depth fixed as endpoints to a chord as in Fig.~\ref{ch6_notch_curv}, we varied the sagitta length on $[-s_{max},0]$ (concave) and $[0,s_{max}]$ (convex) to produce deformed notches while other parameters remained fixed. 
Fig.~\ref{fig:curv_cc_data_notch1} depicts the critical current variation for convex notches. 
Curvature deformity at notches near the shift current source incurs the highest variations and has a significant impact on CC. 
The trend is not linear, but can be approximated as linear for $\leq5\%$ variation, shown in Fig.~\ref{fig:depth_ub_data_notches}.  The change in UB in all cases is indistinguishable from the perfect nanowire value of $7.9\times10^{11} A/m^2$. Similarly, for concave-type deformities, the change in both CC and UB are also indistinguishable. 

From these results a measurable impact on CC and UB currents is demonstrated from all parameters, but the relative weights varied significantly with notch position and curvature having larger impacts than depth and width variation.  We use this data to populate a fault model in the next section. 

\subsection{Modeling Pinning Fault Likelihood}
We used a least squares regression to fit a curve approximating the relationship of critical and upper bound currents as a function of variations in width, depth, and sagitta, respectively.  After adjusting for notch position, these curves and their respective residuals were combined via the total differential method to produce the uncertainties in our nominal critical ($5.1\times10^{11} A/m^2$) and upper bound ($7.9\times10^{11} A/m^2$) currents.  We assume that a device which varies in any of our parameters by greater than $\pm5\%$ is discarded at fabrication time, represented by applying a truncated normal distribution on all parameters.  For curvature, we model only the effects of convex curvature on the critical shift current, in keeping with our results from Fig.~\ref{ch6_curr_var}. To model process variation, each distribution is scaled to match a coefficient of variation of 0.2.  Using the resultant distributions, we can quantify the probability $P_i$ that the applied shift current of $6.5\times10^{11} A/m^2$ will fall above the critical current plus the uncertainty and below the upper bound current minus the uncertainty at any given notch position $i$.  Since all previous domain positions must shift correctly for domain at notch $i$ to shift correctly, we define the probability that a DW at notch position $i$ shifts correctly as $Q_i = \Pi_{n=1}^i P_i$.  Likewise, the cumulative product of $Q_i$ across all notches forms the probability of a correct shift operation, thus we define the probability of pinning as $P_{pin} = 1 - (\Pi_{m=1}^{15} Q_m)$.  Based on our simulation results, this model predicts that faults associated with pinning occur with 
$1.58 \times 10^{-8}$ probability. When there are fewer DWs than notches, this rate could improve depending on wall positions.

\subsection{DWM System Study}
To evaluate the impact of pinning faults, an 8-way 4MB LLC cache and a 8-way 32KB L1 DWM cache architecture~\cite{FusedCache} with our fault model were integrated into and simulated using the Sniper multi-core simulator~\cite{Sniper} presuming four out-of-order cores running at a clock speed of 3 GHz.  DWM nanowire data lengths of $n=32$ and cache lines of 512-bits created DBCs composed of 512*32=16Kbs.  Workloads were 14 benchmarks from SPEC-CPU2006~\cite{spec2006}.

Presuming a $4.55\times10^{-5}$ misalignment fault rate~\cite{hifi}, prior work is able to achieve a mean-time-to-failure (MTTF) of 15 years~\cite{ollivier2019dsn} and 69 years~\cite{hifi}.  When introducing pinning faults at $1.58\times10^{-8}$ probability, the MTTF drops to 2 seconds.
\section{Conclusion}
In this paper, we described the understudied pinning fault in domain wall memory shifting and characterize it with the change in critical shift current due to process variation in notches.   Together with our models of this variation with respect to variation in width, depth, and the new parameter of curvature, we predict that there is a small but non-negligible probability of pinning fault occurrence at $1.58 \times 10^{-8}$, \textcolor{black}{necessitating the development of techniques to mitigate such faults in DWM circuits and designs and to inform circuit level device descriptions such as Verilog-A models.} We observe that the impact on current variation is significant when there are deformities at the notches close to the shift port, which supports targeted mitigation and/or special attention to notch quality at DWM shift points.  

\textcolor{black}{In our future work we will study in detail how pinning can be detected in DWM nanowires.  
Using the information that detecting different DW speeds at the extremity indicates a pinning fault, we plan to develop new shifting fault detection and tolerance approaches for correcting pinning faults in DWM memories.  New fault-tolerance for process variation, quantified by this modeling, is vital to reaping the benefits of DWM in next-generation tiered memory systems.}

\bibliographystyle{ieeetr}
\bibliography{bib/intro,bib/DW,bib/DW3,bib/PF,bib/jones}

\begin{thebibliography}{10}

\bibitem{book94}
L.~H. Diez and D.~Ravelosona, ``Controlling magnetism by interface
  engineering,'' in {\em Magnetic Nano-and Microwires}, pp.~361--379, Elsevier,
  2020.

\bibitem{book95}
J.~Cai, B.~Fang, C.~Wang, and Z.~Zeng, ``Multilevel storage device based on
  domain-wall motion in a magnetic tunnel junction,'' {\em Applied Physics
  Letters}, vol.~111, no.~18, p.~182410, 2017.

\bibitem{book96}
S.~Deb and A.~Chattopadhyay, ``Spintronic device-structure for low-energy xor
  logic using domain wall motion,'' in {\em Proc. of ISCAS}, 2019.

\bibitem{TapeCache}
R.~Venkatesan, V.~Kozhikkottu, C.~Augustine, A.~Raychowdhury, K.~Roy, and
  A.~Raghunathan, ``Tapecache: a high density, energy efficient cache based on
  domain wall memory,'' in {\em Proc. of ISLPED}, pp.~185--190, 2012.

\bibitem{DWM_Tapestri}
R.~Venkatesan, M.~Sharad, K.~Roy, and A.~Raghunathan, ``Dwm-tapestri-an energy
  efficient all-spin cache using domain wall shift based writes,'' in {\em
  Proc. of DATE}, pp.~1825--1830, 2013.

\bibitem{a}
I.~Miron {\em et~al.}, ``Fast current-induced domain-wall motion controlled by
  the rashba effect,'' {\em Nature materials}, vol.~10, no.~6, pp.~419--423,
  2011.

\bibitem{b}
W.-G. Yang and H.~Schmidt, ``Acoustic control of magnetism toward
  energy-efficient applications,'' {\em App. Phys. Rev.}, vol.~8, no.~2, 2021.

\bibitem{hifi}
C.~{Zhang}, G.~{Sun}, X.~{Zhang}, W.~{Zhang}, W.~{Zhao}, T.~{Wang}, Y.~{Liang},
  Y.~{Liu}, Y.~{Wang}, and J.~{Shu}, ``Hi-fi playback: Tolerating position
  errors in shift operations of racetrack memory,'' in {\em Proc. of ISCA},
  2015.

\bibitem{ollivier2019dsn}
S.~Ollivier, D.~Kline~Jr., R.~Kawsher, R.~Melhem, S.~Bhanja, and A.~K. Jones,
  ``Leveraging transverse reads to correct alignment faults in domain wall
  memories,'' in {\em Proc. of DSN}, 2019.

\bibitem{FusedCache}
H.~{Xu}, Y.~{Alkabani}, R.~{Melhem}, and A.~K. {Jones}, ``Fusedcache: A
  naturally inclusive, racetrack memory, dual-level private cache,'' {\em IEEE
  Transactions on Multi-Scale Computing Systems}, vol.~2, no.~2, pp.~69--82,
  2016.

\bibitem{xu2015multilane}
H.~Xu, Y.~Li, R.~Melhem, and A.~K. Jones, ``Multilane racetrack caches:
  Improving efficiency through compression and independent shifting,'' in {\em
  The 20th Asia and South Pacific Design Automation Conference}, pp.~417--422,
  IEEE, 2015.

\bibitem{Castrillon-Polyhedral}
A.~A. Khan, H.~Mewes, T.~Grosser, T.~Hoefler, and J.~Castrillon, ``Polyhedral
  compilation for racetrack memories,'' {\em IEEE TCAD}, vol.~39, no.~11,
  pp.~3968--3980, 2020.

\bibitem{Castrillon-ShiftsReduce}
A.~A. Khan, F.~Hameed, R.~Bl\"{a}sing, S.~S.~P. Parkin, and J.~Castrillon,
  ``Shiftsreduce: Minimizing shifts in racetrack memory 4.0,'' {\em ACM Trans.
  Archit. Code Optim.}, vol.~16, Dec. 2019.

\bibitem{parkin2014dwm}
S.~S.~P. Parkin, L.~Thomas, and S.-H. Yang, ``Method and system for measurement
  of road profile,'' April 2014.
\newblock US Patent 8,687,415 B2.

\bibitem{pinning_Piramanayagam}
M.~Al~Bahri, B.~Borie, T.~Jin, R.~Sbiaa, M.~Kl\"aui, and S.~Piramanayagam,
  ``Staggered magnetic nanowire devices for effective domain-wall pinning in
  racetrack memory,'' {\em Phys. Rev. Applied}, vol.~11, p.~024023, Feb 2019.

\bibitem{thomas2006oscillatory}
L.~Thomas, M.~Hayashi, X.~Jiang, R.~Moriya, C.~Rettner, and S.~S. Parkin,
  ``Oscillatory dependence of current-driven magnetic domain wall motion on
  current pulse length,'' {\em Nature}, vol.~443, no.~7108, 2006.

\bibitem{suzuki2008analysis}
T.~Suzuki, S.~Fukami, N.~Ohshima, K.~Nagahara, and N.~Ishiwata, ``Analysis of
  current-driven domain wall motion from pinning sites in nanostrips with
  perpendicular magnetic anisotropy,'' {\em Journal of Applied Physics},
  vol.~103, no.~11, p.~113913, 2008.

\bibitem{hayashi2007current}
M.~Hayashi, {\em Current driven dynamics of magnetic domain walls in permalloy
  nanowires}.
\newblock Stanford University California, 2007.

\bibitem{sampaio2015domain}
J.~Sampaio, J.~Grollier, and P.~Metaxas, ``Domain wall motion in
  nanostructures,'' in {\em Hand. of Surface Science}, vol.~5, pp.~335--370,
  Elsevier Science, 2016.

\bibitem{aharoni2000introduction}
A.~Aharoni {\em et~al.}, {\em Introduction to the Theory of Ferromagnetism},
  vol.~109.
\newblock Clarendon Press, 2000.

\bibitem{berger1984exchange}
L.~Berger, ``Exchange interaction between ferromagnetic domain wall and
  electric current in very thin metallic films,'' {\em Journal of Applied
  Physics}, vol.~55, no.~6, pp.~1954--1956, 1984.

\bibitem{tatara2004theory}
G.~Tatara and H.~Kohno, ``Theory of current-driven domain wall motion: Spin
  transfer versus momentum transfer,'' {\em Physical review letters}, vol.~92,
  no.~8, p.~086601, 2004.

\bibitem{tatara2008microscopic}
G.~Tatara, H.~Kohno, and J.~Shibata, ``Microscopic approach to current-driven
  domain wall dynamics,'' {\em Physics Reports}, vol.~468, no.~6, 2008.

\bibitem{iyengar2014modeling}
A.~Iyengar and S.~Ghosh, ``Modeling and analysis of domain wall dynamics for
  robust and low-power embedded memory,'' in {\em DAC}, 2014.

\bibitem{mumax}
A.~Vansteenkiste, J.~Leliaert, M.~Dvornik, M.~Helsen, F.~Garcia-Sanchez, and
  B.~Van~Waeyenberge, ``The design and verification of mumax3,'' {\em AIP
  Advances}, vol.~4, no.~10, p.~107133, 2014.

\bibitem{Sniper}
T.~E. Carlson, W.~Heirman, S.~Eyerman, I.~Hur, and L.~Eeckhout, ``An evaluation
  of high-level mechanistic core models,'' {\em ACM Trans. Archit. Code
  Optim.}, vol.~11, aug 2014.

\bibitem{spec2006}
J.~L. Henning, ``Spec cpu2006 benchmark descriptions,'' {\em ACM SIGARCH
  Computer Architecture News}, vol.~34, pp.~1--17, sept 2006.

\end{thebibliography}

\end{document}